\documentclass[11pt,nofootinbib]{article}
\pdfoutput=1
\usepackage{jheppub}
\usepackage{graphicx}
\usepackage{slashed}
\usepackage[normalem]{ulem}
\usepackage{wasysym}
\usepackage{verbatim}
\usepackage{cancel}
\usepackage{natbib}
\usepackage{amsmath,amssymb,amsfonts}
\usepackage{enumerate}
\usepackage{subfigure} 
\usepackage{caption,subcaption}

\newcommand\beq{\begin{equation}}
\newcommand\eeq{\end{equation}}

\newcommand\hc{\text{h.c.}}

\title{Novel probes for electron-muon flavor violation from exotic Higgs decays}
\author[a]{P. Uttayarat,}
\author[b]{J. Julio}
\author[c]{and R. Primulando}

\affiliation[a]{Department of Physics, Srinakharinwirot University, 114 Sukhumvit 23rd Rd., Wattana, Bangkok 10110, Thailand}
\affiliation[b]{National Research and Innovation Agency, KST B.\,J.\,Habibie, South Tangerang 15314, Indonesia}
\affiliation[c]{Center for Theoretical Physics, Department of Physics, Parahyangan Catholic University, Jl. Ciumbuleuit 94, Bandung 40141, Indonesia}

\emailAdd{patipan@g.swu.ac.th}
\emailAdd{julio@brin.go.id}
\emailAdd{rprimulando@unpar.ac.id}

\abstract{In this paper, we propose two novel signatures of Higgs decays to search for electron–muon flavor violation. These signatures arise from the presence of a light pseudoscalar into which the 125-GeV Higgs boson decays. The pseudoscalar subsequently decays into an electron–muon pair, leading to multilepton final states, which are relatively clean signatures to search for at the LHC. As a benchmark, we consider the type-III Two-Higgs-doublet-model. We analyze both low-energy and collider constraints on the model and identify regions of parameter space where the light pseudoscalar is viable. Our proposed signatures yield stronger constraints on the lepton flavor violating couplings than current low-energy precision measurements. Taken together, our findings suggest that collider-based probes of exotic Higgs decays provide a powerful complement to precision experiments in the quest to uncover new physics.}

\begin{document}
\maketitle

\flushbottom
%%%%%%%%%%%%%%%%%%%%%%%%%%%%%%%%%%%%%%%%%
\section{Introduction}
%%%%%%%%%%%%%%%%%%%%%%%%%%%%%%%%%%%%%%%%
After a decade since the discovery of the Higgs boson, the era of precision Higgs measurements has opened up. The primary goal of these precise measurements is to determine if the Higgs sector strictly adheres to the Standard Model (SM) predictions, or harbors the sign of new physics. Although current measurements are in good agreement with the SM~\cite{ATLAS:2016neq,CMS:2022dwd,ATLAS:2022vkf}, there are still room for small deviations that could point to extensions of the SM. Two major avenues are being pursued to probe such deviations: first, the high precision measurement of the SM-like Higgs couplings to the SM particles; and second, the search of exotic Higgs decays that are not predicted in the SM. 

Among the exotic decays, the lepton-flavor violation (LFV) varieties offer particularly sensitive windows to the new physics. The SM strictly forbids LFV, so any observations of such a decay would be a clear indication of physics beyond the SM. Motivated by this, ATLAS and CMS collaborations have searched for the two-body final states LFV decays, i.e., $h \to \tau^\pm \mu^\mp$ and $h \to \tau^\pm e^\mp$~\cite{CMS:2021rsq,ATLAS:2023mvd}, as well as $h \to \mu^\pm e^\mp$~\cite{ATLAS:2019old,CMS:2023pte}. These searches so far have yielded null results, and hence they place the upper limits on the LFV branching ratios of the Higgs boson of the order 0.1\% for the first two channels and $\mathcal O \left(10^{-5}\right)$ for the latter mode.

These collider bounds must be interpreted in light of stringent constraints from the low-energy LFV observables. In the $e\mu$ case, for instance, the low-energy LFV processes $\mu \to e \gamma$ and $\mu \to e$ conversions in nuclei provide extremely tight limits, which are few orders of magnitude stronger than the bounds from $h \to \mu^\pm e^\mp$. Typically, the bounds are derived in the effective field theory (EFT) framework under the assumption that no new degrees of freedom exist around the electroweak scale~\cite{Blankenburg:2012ex,Harnik:2012pb}. However, such an assumption is invalid in general. In fact, the presence of new degrees of freedom around the electroweak scale can significantly affect both the low-energy and the high-energy observables~\cite{Buschmann:2016uzg,Primulando:2023ugc,Koivunen:2023led,Afik:2023vyl,Crivellin:2013wna}. Thus, a consistent interpretation of LFV signals requires embedding the EFT into an ultraviolet (UV)-complete model.

In this work, we explore the LFV Higgs decays within the framework of the type-III Two-Higgs-Doublet-Model (2HDM), which extends the scalar sector by the introduction of the second Higgs doublet without imposing any discrete symmetries for flavor conservations. This allows for a tree-level LFV Higgs couplings. Unlike effective theories, the 2HDM provides a concrete UV structure linking LFV observables across energy scales. Importantly, it opens up additional decay channels for the SM-like Higgs bosons through its interactions with other scalar states.

We consider a specific scenario in which the pseudoscalar $A$ is lighter than the SM-like Higgs boson $h$. In this setup, the $h$ can decay via $h \to A Z$ and $h \to AA$, with the pseudoscalar subsequently decaying through the LFV channel $A \to \mu^\pm e^\mp$. These processes lead to distinctive multilepton signatures, such as $h \to \mu^\pm e^\mp \ell^+ \ell^-$ or $h \to \mu^\pm e^\mp \mu^\pm e^\mp$, with little background from SM processes. Although no LHC searches have yet targeted these exact final states, existing analyses in the related multilepton or exotic Higgs channels can be repurposed to constrain them.

This paper presents a systematic study of these novel LFV signatures at the LHC. In section~\ref{sec:model}, we provide a brief summary of the type-III 2HDM. We then discuss its low-energy LFV observables in section~\ref{sec:lfv}. In section~\ref{sec:colliderbound}, we discuss the relevant collider constraints for the scenario where the pseudoscalar is lighter than the SM-like Higgs boson. We then propose novel collider signatures for the SM-like Higgs decays in section~\ref{sec:collider}. We then conclude and discuss our results in section~\ref{sec:conc}.

%%%%%%%%%%%%%%%%%%%%%%%%%%%%%%%%%%%%%%%%%
\section{Type-III 2HDM}
\label{sec:model}
%%%%%%%%%%%%%%%%%%%%%%%%%%%%%%%%%%%%%%%%

In this section, we give a brief overview of the Type-III 2HDM. We will closely follow the notation and convention of ref.~\cite{Primulando:2016eod}. The two electroweak scalar doublets are denoted by $\Phi_1$ and $\Phi_2$. Their scalar potential is given by
\begin{equation}
\begin{aligned}
	V =&~ M_{11}^2 (\Phi_1^{\dagger} \Phi_1^{\phantom{\dagger}}) + M_{22}^2 (\Phi_2^{\dagger} \Phi_2^{\phantom{\dagger}}) -  [M_{12}^2(\Phi_1^{\dagger} \Phi_2^{\phantom{\dagger}}) + \text{h.c.}]  \\
&+ \frac{1}{2} \lambda_1 (\Phi_1^{\dagger} \Phi_1^{\phantom{\dagger}})^2 + \frac{1}{2} \lambda_2 (\Phi_2^{\dagger} \Phi_2^{\phantom{\dagger}})^2 + \lambda_3 (\Phi_1^{\dagger} \Phi_1^{\phantom{\dagger}})(\Phi_2^{\dagger} \Phi_2^{\phantom{\dagger}}) + \lambda_4 (\Phi_1^{\dagger} \Phi_2^{\phantom{\dagger}})(\Phi_2^{\dagger} \Phi_1^{\phantom{\dagger}}) \\
&+ \left\{\frac{1}{2} \lambda_5 (\Phi_1^{\dagger} \Phi_2^{\phantom{\dagger}})^2 + \left[\lambda_6 (\Phi_1^{\dagger} \Phi_1^{\phantom{\dagger}}) + \lambda_7 (\Phi_2^{\dagger} \Phi_2^{\phantom{\dagger}})\right] (\Phi_1^{\dagger} \Phi_2^{\phantom{\dagger}}) + \text{h.c.} \right\}.
\end{aligned}
\end{equation}
In principle, the parameters $M_{12}^2$, $\lambda_5$, $\lambda_6$ and $\lambda_7$ can be complex, which would lead to $CP$-violation in the scalar sector. In this work, for simplicity, we will assume that such parameters are real, so that the scalar sector is $CP$ symmetric. For the present discussion, it will be more convenient to work in the Higgs basis~\cite{Georgi:1978ri}, that is, the doublets $\Phi_1$ and $\Phi_2$ are expanded as
\begin{equation}
	\Phi_1 = \begin{pmatrix}G^+\\ \frac{v+\phi_1+iG^0}{\sqrt{2}}\end{pmatrix},
	\qquad
	\Phi_2 = \begin{pmatrix}H^+\\ \frac{\phi_2+iA}{\sqrt{2}}\end{pmatrix},
\end{equation}
where $v=246$ GeV is the electroweak vacuum expectation value, $G^+$ and $G^0$ are the would-be Goldstone bosons, $H^+$ is the physical charged Higgs boson, $A$ is the physical $CP$-odd Higgs boson and $\phi_{1,2}$ are two $CP$-even neutral scalars. 

From the minimization of the scalar potential, one can express some parameters in terms of the others. For instance,
\begin{align}
    M_{11}^2 = -\tfrac{1}{2}\lambda_1v^2, \quad M_{12}^2 = \tfrac{1}{2}\lambda_6v^2.
\end{align}
From here, one can derive the masses of $H^\pm$ and $A$ 
\begin{align}
	m_{H^\pm}^2 &= M_{22}^2 + \frac{\lambda_3}{2}v^2 , \label{eq:mHp}\\
	m_A^2 &= m_{H^\pm}^2 + \frac{\lambda_4 - \lambda_5}{2} v^2,
    \label{eq:mA} 
\end{align}
and the mass matrix of $\phi_1$ and $\phi_2$ in the basis of $(\phi_1,\phi_2)$
\begin{align}
    \mathcal{M}^2 = \begin{pmatrix}
        \lambda_1 v^2 & \lambda_6 v^2 \\ \lambda_6 v^2 & m^2_{H^\pm} + \tfrac{1}{2}(\lambda_4+\lambda_5) v^2
    \end{pmatrix}.
\end{align}
The above mass matrix is diagonalized by rotating the basis into $h$ and $H$ mass eigenbasis
\begin{equation}
	\begin{pmatrix}\phi_1\\ \phi_2\end{pmatrix}= \begin{pmatrix} \phantom{-}c_\alpha & s_\alpha \\ -s_\alpha & c_\alpha\end{pmatrix}
	\begin{pmatrix}h\\ H \end{pmatrix},
\end{equation}
where $c_x,s_x$ denote $\cos x,\sin x$. The mixing angle is given by
\begin{equation}
	%\tan2\alpha = \frac{2\lambda_6v^2}{m_A^2-(\lambda_1-\lambda_5)v^2}, \quad \text{or} \quad
    s_{2\alpha} = \frac{2\lambda_6 v^2}{m_H^2-m_h^2},
\end{equation}
resulting in eigenvalues 
\begin{equation}
	m_{h,H}^2 = \frac{1}{4}\left( 2m_{H^\pm}^2 + (2\lambda_1 + \lambda_4 + \lambda_5) v^2 \mp \sqrt{\left( 2m_{H^\pm}^2 + (-2\lambda_1 +\lambda_4+ \lambda_5) v^2\right)^2 + 16\lambda_6^2 v^4 } \right).
\end{equation}
In this paper, we identify $h$ with the 125-GeV Higgs boson and $H$ with the heavy $CP$-even Higgs boson. The bound on the mixing angle $\alpha$ is  determined from the combined Run 1 and Run 2 measurements of the ATLAS and CMS, which yields $s_\alpha\le0.15$ at 95\% confidence level (CL)~\cite{ATLAS:2016neq,CMS:2022dwd,ATLAS:2022vkf}. In the small $s_\alpha$ limit, the masses of $h$ and $H$ are approximated as
\begin{align}
    m_h^2 &= \lambda_1v^2 + \mathcal{O}(s^2_\alpha),\\
    m_H^2 &= m_{H^\pm}^2 + \frac{1}{2}(\lambda_4 + \lambda_5) v^2 + \mathcal{O}(s^2_\alpha).
    \label{eq:sc-neutral}
\end{align}

The presence of $M_{22}^2$ makes it possible for $H$, $A$ and $H^\pm$ masses to be much greater than the weak scale. However, their mass-squared differences are controlled by quartic couplings, $\lambda_i$, theoretically constrained by perturbativity, vacuum stability and unitarity conditions. In our analysis, we take $|\lambda_i|<4\pi$ for perturbativity. The vacuum stability and unitarity constraints, in the presence of $\lambda_6$ and $\lambda_7$, are complicated and not illuminating. The coupling $\lambda_6$ is directly proportional to the mixing angle $s_\alpha$, so it is expected to be small. The coupling $\lambda_7$, on the other hand, does not play any role in our analysis. Hence, for simplicity, we will take $\lambda_7 = 0$. With these simplifications, the vacuum stability constraints read~\cite{Ferreira:2004yd,Ivanov:2006yq,Ferreira:2009jb}\footnote{The constraint involving $\lambda_6$ is only the necessary condition for the scalar potential to be bounded below.}

\begin{align}
    \lambda_1,\,\lambda_2 &>  0, \\
    \lambda_3 &> -\sqrt{\lambda_1\lambda_2}\,,\\
    \lambda_3 + \lambda_4 - |\lambda_5| &> -\sqrt{\lambda_1\lambda_2}\,,
    \label{eq:stability345} \\
    \lambda_1+\lambda_2+2(\lambda_3+\lambda_4+\lambda_5) &> 4|\lambda_6|,
    %\lambda_1+\lambda_2+2(\lambda_3+\lambda_4+\lambda_5) &> 4|\lambda_6+\lambda_7|.
    \label{eq:vacuumstability}
\end{align}
and the unitarity constraints are given by~\cite{Ginzburg:2003fe}
\begin{align}
    3(\lambda_1+\lambda_2) + \sqrt{9(\lambda_1-\lambda_2)^2+(2\lambda_3+\lambda_4)^2} &< 16\pi,\\
    \lambda_1+\lambda_2 + \sqrt{(\lambda_1-\lambda_2)^2+4\lambda_4^2} &< 16\pi,\\
    \lambda_1+\lambda_2 + \sqrt{(\lambda_1-\lambda_2)^2+4\lambda_5^2} &< 16\pi,\\
    |\lambda_3 + 2\lambda_4 \pm 3\lambda_5| &< 8\pi,\\
    |\lambda_3\pm\lambda_4| &< 8\pi,\\
    |\lambda_3\pm\lambda_5| &< 8\pi.
    \label{eq:unitarity}
\end{align}

The scalar mass splittings are also constrained by the $T$ parameter, inferred from the electroweak precision measurements. The new physics correction to the $T$ parameter reads
\begin{align}
    \Delta T &= \frac{1}{16\pi^2\alpha_{em}v^2}\left[F(m_{H^\pm}^2,m_H^2)+F(m_{H^\pm}^2,m_A^2)-F(m_A^2,m_H^2)+\mathcal{O}(s^2_\alpha)\right],
    %&\quad+\left.\sin^2\alpha\left[F(m_{H^\pm}^2,m_h^2)-F(m_{H^\pm}^2,m_H^2)+F(m_{H}^2,m_A^2)-F(m_{h}^2,m_A^2)\right]\right\},
\end{align}
where $\alpha_{em}^{-1}=137$ is the electromagnetic fine-structure constant in the Thompson limit, whereas $F(x,y)$ is defined as
\begin{align}
    F(x,y) = \frac{x+y}{2}- \frac{xy}{x-y}\ln\frac{x}{y}.
\end{align}
One can notice that if one of the neutral scalar masses is degenerate with the charged one, $\Delta T$ will vanish. This is consistent with the value of $\Delta T$ inferred from the fit to the electroweak precision data, which is $\Delta T=0.00\pm0.06$~\cite{ParticleDataGroup:2024cfk}. 

In this work, we shall consider the case with $m_A\lesssim 100~\text{GeV}$ with heavier $m_H$ and $m_{H^\pm}$. To comply with the $T$ parameter constraint, it is instructive to assume $m_H=m_{H^\pm}$, which in turn implies $\lambda_4+\lambda_5=0$, see eq.~\eqref{eq:sc-neutral}. The value of $m_{H^\pm}$ is constrained by LHC searches. For leptophilic charged scalar decaying 100\% into electron or muon, it is constrained to be heavier than 550 GeV. As a consequence, by virtue of equations.~\eqref{eq:mA} and \eqref{eq:stability345}, we expect large quartic couplings. We will discuss it in detail in section~\ref{sec:collider}. 

The Higgs doublet $\Phi_1$ is responsible to generate fermion masses thorugh Yukawa interactions
\begin{equation}
    \mathcal{L}_{Yuk} \supset  -\frac{\sqrt{2}m^i_\ell}{v}\delta^{ij}\bar{L}_i\ell_{Rj}\Phi_1 
    -\frac{\sqrt{2}m^i_U}{v}\delta^{ij}\bar{Q}_iu_{Rj}\tilde\Phi_1
    -\frac{\sqrt{2}m^k_D}{v}V^{ik}\delta^{kj}\bar{Q}_id_{Rj}\Phi_1 + \hc,
\end{equation}
where $m_\ell$, $m_U$ and $m_D$ are the diagonal lepton, up-type quark and down-type quark mass matrices respectively, $V$ is the Cabibbo-Kobayashi-Maskawa matrix and $\tilde\Phi_a\equiv i\sigma_2\Phi^\ast_a$. In the above equation, the left-handed fermion doublets are taken to be
\begin{equation}
    L = \begin{pmatrix}\nu_L \\ \ell_L\end{pmatrix},\qquad
	Q = \begin{pmatrix}u_L \\ V d_L\end{pmatrix},
\end{equation}
where fields $\ell_{L(R)}$, $u_{L(R)}$ and $d_{L(R)}$ are defined in their respected mass eigenbases. The Yukawa couplings of $\Phi_2$ in general lead to flavor violation. In this work, we will focus on the leptonic couplings
\begin{equation}
    \mathcal{L}_{Yuk} \supset -\sqrt{2}Y_{ij}\bar{L}_i\ell_{Rj} \Phi_2 +\hc.
    \label{eq:lfv}
\end{equation}
The coupling $Y_{ij}$ in principle can be complex, but for simplicity, we assume that they are real. In our analysis, we mainly focus on the $Y_{e\mu}$ and $Y_{\mu e}$ couplings. However, other components of the $Y$ matrix, e.g., $Y_{\tau\tau}$, can also be present. This observation will play an important role when considering the collider searches in multilepton channels to be discussed in section~\ref{sec:collider}.

%%%%%%%%%%%%%%%%%%%%%%%%%%%%%%%%%%%%%%%%%
\section{Low-energy lepton flavor-violating constraints}
\label{sec:lfv}
%%%%%%%%%%%%%%%%%%%%%%%%%%%%%%%%%%%%%%%%

The Yukawa couplings in eq.~\eqref{eq:lfv}, together with the neutral scalar mixing angle $s_\alpha$, lead to LFV decays of charged lepton. In particular, nonzero $Y_{e\mu}$ and $Y_{\mu e}$ give rise to $\mu\to e\gamma$ and $\mu\to e$ conversion in atomic nuclei. These low-energy LFV processes can be described by effective operators
\begin{align}
   \mathcal{L}_{eff} =&~  -\frac{em_\mu}{16\pi^2v^2}\bar{e}\sigma^{\rho\lambda}c_LP_L\mu F_{\rho\lambda} -\frac{1}{2v^2}\sum_q\left[g_{LV}^q(\bar e \gamma^\rho P_L\mu)(\bar q\gamma_\rho q) + g_{LS}^q(\bar e P_R\mu)(\bar q q)\right] \nonumber \\
   &+ (L\leftrightarrow R)+\hc,
    \label{eq:lowenergylfv}
\end{align}
where $P_{R,L}=(1\pm\gamma_5)/2$ are the right- and left-chirality projection operators, respectively. 
Terms with coefficients $c_{L,R}$ are dipole operators, which are relevant for both $\mu\to e\gamma$ and $\mu\to e$ conversion processes. The last two terms are effective vector and scalar current interactions between leptons and quarks which are relevant for $\mu\to e$ conversion.

\begin{figure}
    \centering
    \includegraphics[width=\columnwidth]{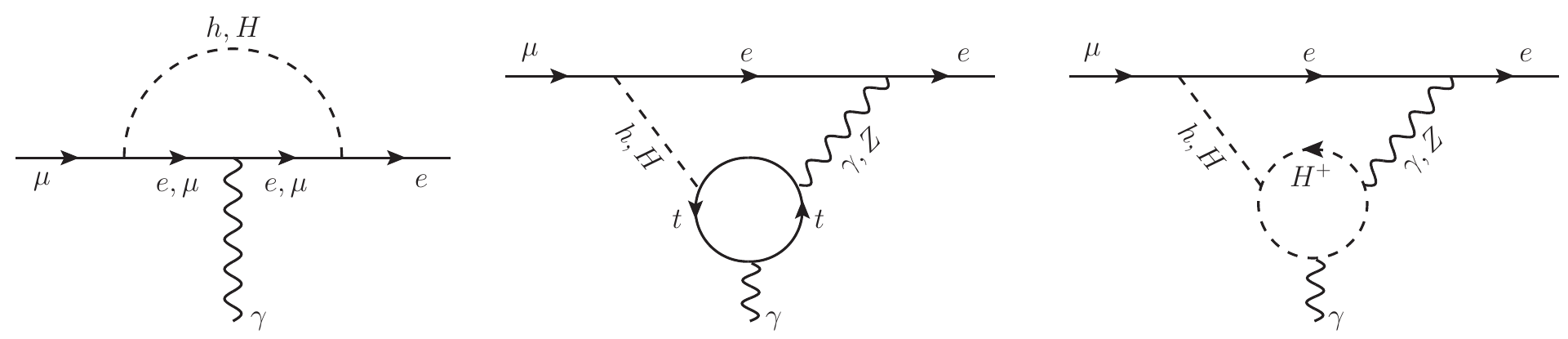}
    \caption{Representative Feynman diagrams for $\mu\to e\gamma$ decay at one- and two-loop levels. The middle diagram is the representation of Barr-Zee diagram, while the rightmost diagram is the example of diagrams induced by the quartic couplings $\lambda_i$.}
    \label{fig:muedecay}
\end{figure}

The partial decay width for $\mu\to e \gamma$ is given  by
\begin{equation}
    \Gamma_{\mu\to e \gamma} = \frac{\alpha_{em}m_\mu^5}{256\pi^4v^4}\left(|c_L|^2+|c_R|^2\right).
\end{equation}
The coefficients $c_{L,R}$ are induced by quantum effects and are expressed as $c_{L,R}=c_{L,R}^{(1)}+c_{L,R}^{(2)}$, with the superscript denoting the loop level. The one-loop contributions are induced through Feynman diagram in  figure~\ref{fig:muedecay} (left) and are given by~\cite{Hisano:1995cp}
\begin{align}
    c_L^{(1)} =&~ -\frac{s_{2\alpha}}{6}m_\mu vY_{\mu e}\left[\frac{1}{m_h^2}\left(4 + 3\ln\frac{m_\mu^2}{m_h^2}\right) - \frac{1}{m_H^2}\left(4 + 3\ln\frac{m_\mu^2}{m_H^2}\right) \right]. \nonumber \\
    c_R^{(1)} =&~ c_L^{(1)}\bigg|_{Y_{\mu e}\to Y_{e\mu}}.
\end{align}

It has long been realized in the literature that numerically large contributions to the dipole operators can arise from two-loop Feynman diagrams due to possible large hierarchies among the couplings~\cite{Weinberg:1989dx,Dicus:1989va,Barr:1990vd}. For example, the so-called Barr-Zee diagram with an internal top-quark loop, shown in the middle of  figure~\ref{fig:muedecay}, results in~\cite{Chang:1993kw,Davidson:2010xv,Crivellin:2014cta} 
\begin{equation}
    %c_L^{(2t)} = -\frac{2s_{2\alpha}}{3}\frac{\alpha_{em}Y_{\mu e}}{\pi}\frac{m_t}{m_\mu}\left[\frac{m_tv}{m_h^2}f_2\left(\frac{m_t^2}{m_h^2}\right) - \frac{m_tv}{m_H^2}f_2\left(\frac{m_t^2}{m_H^2}\right) \right],
    c_L^{(2t)} = \frac{s_{2\alpha}}{3}\frac{\alpha_{em}Y_{\mu e}}{\pi}\frac{m_t}{m_\mu}\left[\frac{m_tv}{m_h^2}f\left(\frac{m_t^2}{m_h^2}\right) - \frac{m_tv}{m_H^2}f\left(\frac{m_t^2}{m_H^2}\right) \right] ,
\end{equation}
where 
\begin{align}
 	f(z) &= \int_0^1dx\frac{1 - 2x(1-x)}{x(1-x) - z}\ln\frac{x(1-x)}{z}\label{eq:loopf}.
\end{align}
For $m_H$ below a TeV scale, the loop function $f(z)\lesssim 4\pi$. Thus, the one-loop coefficient is parametrically suppressed by $\,m_{\mu}^2/(\alpha_{em}m_t^2)$ compared to the two-loop one.
A more complete analysis of two-loop contributions to the dipole coefficients has been carried out in ref.~\cite{Altmannshofer:2025nsl}, including the contributions from the scalar quartic couplings, see figure~\ref{fig:muedecay} (right). Such contributions could be significant, particularly in our scenario where we have large scalar mass splittings. We use the Python code provided in ref.~\cite{Altmannshofer:2025nsl} to compute the coefficients $c_L$ and $c_R$ numerically. 

Experimentally, the most stringent constraint on $\mu\to e\gamma$ decay is provided by the MEG II experiment, with BR$(\mu\to e\gamma)\le1.5\times10^{-13}$ at 90\% CL~\cite{MEGII:2025gzr}. It should be noted that the MEG II experiment is still running, collecting more data. It is projected that by 2026, the MEG II experiment should be able to push the bound down to BR$(\mu\to e\gamma)\le6.0\times10^{-14}$ at 90\% CL.

The $\mu\to e$ conversion rate in atomic nuclei is computed by matching the quark-level effective operators in eq.~\eqref{eq:lowenergylfv} onto the nucleon-level effective operators. The matching is done by evaluating the nucleon matrix elements
\begin{align}
    \langle N|\bar q\gamma_\rho q|N\rangle = n^{(q,N)}\bar N\gamma_\rho N,\\
    \langle N|\bar q q|N\rangle = f^{(q,N)}\frac{m_N}{m_q}\bar NN,
\end{align}
where $n^{(q,N)}$ is the number of valence quark $q$ inside the nucleon $N$ and $f^{(q,N)}$ is the form factor. For light quarks, $q = u,d,s$, the form factors are determined through lattice calculations. The heavy quarks ($q=c,b,t$) form factors are related to those of the light quarks through the trace anomaly~\cite{Shifman:1978zn,Jungman:1995df}. The values of quark form factors are listed in table~\ref{tab:formfactor}.

Equipped with the quark form factors, one can readily compute the $\mu\to e$ conversion rate. It is given by
\begin{equation}
    \Gamma(\mu\to e\text{ conv.}) = m_\mu^5\left|\frac{e\,c_R\,D}{32v^2\pi^2} + (2g_{LV}^u+g_{LV}^d)V^p + \sum_q\sum_{N=p,n}g_{LS}^q\frac{m_N}{m_q}f^{(q,N)}S^N\right|^2 + (L\leftrightarrow R),
\end{equation}
%}
where $D$, $V^p$ and $S^{p,n}$ are the overlap integrals. Their numerical values, for various atomic nuclei, have been tabulated in ref.~\cite{Kitano:2002mt}. The effective couplings $g_{LV}^q$ and $g_{LS}^q$ are given by
\begin{align}
    g_{LV}^q &= \frac{s_{2\alpha}\alpha_{em}Q_q}{36\pi}\frac{m_\mu}{v}Y_{e\mu}\left[\frac{1}{m_h^2}\left(4+3\ln\frac{m_\mu^2}{m_h^2}\right)-\frac{1}{m_H^2}\left(4+3\ln\frac{m_\mu^2}{m_H^2}\right)\right],\\
    g^q_{LS} &= s_{2\alpha}\frac{m_q}{v}Y_{e\mu}\left(\frac{1}{m_h^2}-\frac{1}{m_H^2}\right).
\end{align}
The effective coupling $g_{RV}^q$ and $g_{RS}^q$ can be obtained from their left-handed counterparts by the replacement $Y_{e\mu}\to Y_{\mu e}$. Note that we only include photon-penguin diagrams in vector couplings and tree-level scalar exchange in scalar ones. In principle, there are other contributions from $Z$-penguin and box diagrams, but they are deemed insignificant. The $Z$-exchange diagrams, for example, are suppressed by $m_\mu^2/m_Z^2$, while the box diagrams are suppressed by light quark masses. In addition, the box contributions can also result in scalar operators. However, that requires a chirality flip from muon propagator, inducing an additional $m_\mu/v$ suppression.

\begin{table}[t!]
\begin{center} 
\begin{tabular}{c|c|c|c|c}
\hline
Nucleon & $f^{(u,N)}$~\cite{Bishara:2015cha} & $f^{(d,N)}$~\cite{Bishara:2015cha} &$f^{(s,N)}$~\cite{Junnarkar:2013ac} &$f^{(Q,N)}$~\cite{Harnik:2012pb} \\\hline
$p$ & 0.018$\pm$0.005& 0.034$\pm$0.011 & 0.043$\pm$0.011& 0.067$\pm$0.001\\%\hline
$n$ & 0.016$\pm$0.005& 0.038$\pm$0.011 &0.043$\pm$0.011 & 0.067$\pm$0.001 \\ \hline
\end{tabular}
\caption{Numerical values for $f^{(q,N)}$. Note that $Q$ stands for heavy quarks, $c,b,t$.}
\label{tab:formfactor}
\end{center} 
\end{table}

The experimental search $\mu\to e$ conversion in gold nuclei provides the strongest constraint. The SINDRUM II collaboration has placed an upper limit $\Gamma(\mu\to e$ conv.)/$\Gamma$(captured) $< 7\times10^{-13}$ at 90\% CL~\cite{Bertl:2006up}. The corresponding muon capture rate in the gold nucleus is $13.07\times10^{6}\text{ s}^{-1}$~\cite{Suzuki:1987jf}. Finally, the overlap integrals for the gold nucleus are given by $D = 0.189$, $V^p = 0.0974$, $S^p=0.0614$ and $S^n = 0.0918$.

\begin{figure}[t]
\begin{center}
\includegraphics[width=0.5\textwidth]{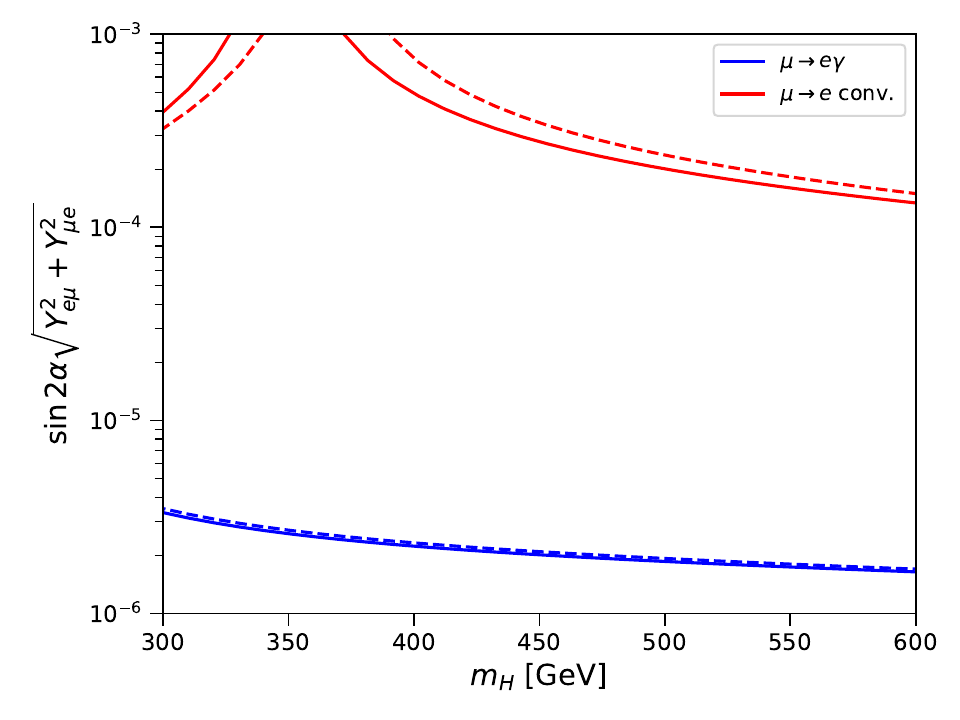}
\caption{Conservative upper bounds on $s_{2\alpha}\sqrt{Y_{e\mu}^2 + Y_{\mu e}^2}$ from low energy LFV searches as a function of $m_H=m_{H^\pm}$. The solid (dashed) lines show the constraints when $h\to AA$ is open (closed). In in case where the decay $h\to AA$ is closed, the quartic coupling $\lambda_3$ is taken to be at the minimum allowed by vacuum stability constraint so that the upper limits are the least constraining.
}
\label{fig:lfvconstraint}
\end{center}
\end{figure}

As has already been mentioned, the $\mu\to e\gamma$ and $\mu\to e$ conversion constraints depend on the quartic couplings $\lambda_3$, $\lambda_4$ and $\lambda_5$ through the dipole coefficients $c_L$ and $c_R$. The couplings $\lambda_4$ and $\lambda_5$ are determined from the scalar masses $m_A$, $m_H$ and $m_{H^\pm}$, leaving $\lambda_3$ as an independent parameter. However, in the event that $h\to AA$ is open, the ATLAS $h\to AA$ search dictates $\lambda_3\simeq 2(m_{H^\pm}^2-m_A^2)/v^2$, see section~\ref{subsec:haa}. 

In a scenario where $m_A\le100$ GeV and $m_H=m_{H^\pm}\gtrsim 300$ GeV, we find that the $\mu\to e\gamma$ and $\mu\to e$ conversion constraints depend only on whether the decay channel $h\to AA$ is open or closed, and not on the actual value of $m_A$. In this region of parameter space, $\mu\to e\gamma$ provides a tighter constraint than $\mu\to e$ conversion. Figure~\ref{fig:lfvconstraint} shows the conservative bounds on $s_{2\alpha}\sqrt{Y_{e\mu}^2+Y_{\mu e}^2}$ as a function of $m_H$ for the cases where the decay channel $h\to AA$ is open (solid lines) and closed (dashed lines). In obtaining these bounds in the scenario where $h\to AA$ is closed, the coupling $\lambda_3$ is taken to be at the minimum value allowed by Eq.~\eqref{eq:stability345}, so that the constraint from $\mu\to e\gamma$ is at its weakest. 
%The bounds for other values of $\lambda_3$ satisfying vacuum stability constraint roughly lie between the solid and dashed lines.
Allowing $\lambda_3$ to take on a larger value improves the $\mu\to e\gamma$ limit by at most 25\%  (4\%) for $m_H$ = 300 (600) GeV.

%%%%%%%%%%%%%%%%%%%%%%%%%%%%%%%%%%%%%%%%
\section{Current collider constraints on the model}
\label{sec:colliderbound}
%%%%%%%%%%%%%%%%%%%%%%%%%%%%%%%%%%%%%%%%
In our setup, the doublet $\Phi_2$ couples to leptons but not quarks, see eq.~\eqref{eq:lfv}. As a result, only the heavy scalar $H$ can be singly produced, provided that it mixes with the SM-like Higgs boson. In our analysis, we will focus on the LFV decays induced by coupling $Y_{e\mu}$ and $Y_{\mu e}$. The CMS collaboration has conducted a dedicated search for the heavy Higgs boson decaying into an electron-muon pair in the mass range 110 GeV $< m_H <$ 160 GeV~\cite{CMS:2023pte}. This search provides the most stringent bound on the LFV decay of $H$. For heavier $m_H$, the LFV constraints can be obtained from searches for a resonant production of the sneutrino decaying to $e\mu$. The mass window of 160--200 GeV is covered by the D0 search~\cite{D0:2010rhn}. The corresponding CMS search covers the mass from 200 GeV up to several TeV~\cite{CMS:2022fsw}. We reinterpret these analyses by assuming that $H$ and the sneutrino have comparable signal acceptance. Since the $H$ production cross-section is proportional to the mixing angle squared, the resulting limits are presented in terms of $s^2_\alpha \times$ BR$(H \rightarrow e^\pm\mu^\mp)$, see figure~\ref{fig:H}. Note that the LFV constraints of the D0 search are weak, with $s^2_\alpha \times$ BR$(H \rightarrow e^\pm\mu^\mp) < \mathcal{O}\left(10^{-2}\right)$. We further note that for $m_H > m_A + m_Z$, the decay $H \rightarrow A Z$ dominates due to the smallness of the LFV Yukawas, suppressing the branching ratio BR$(H \rightarrow e^\pm\mu^\mp)$.

\begin{figure}[t]
\begin{center}
\includegraphics[width=0.5\textwidth]{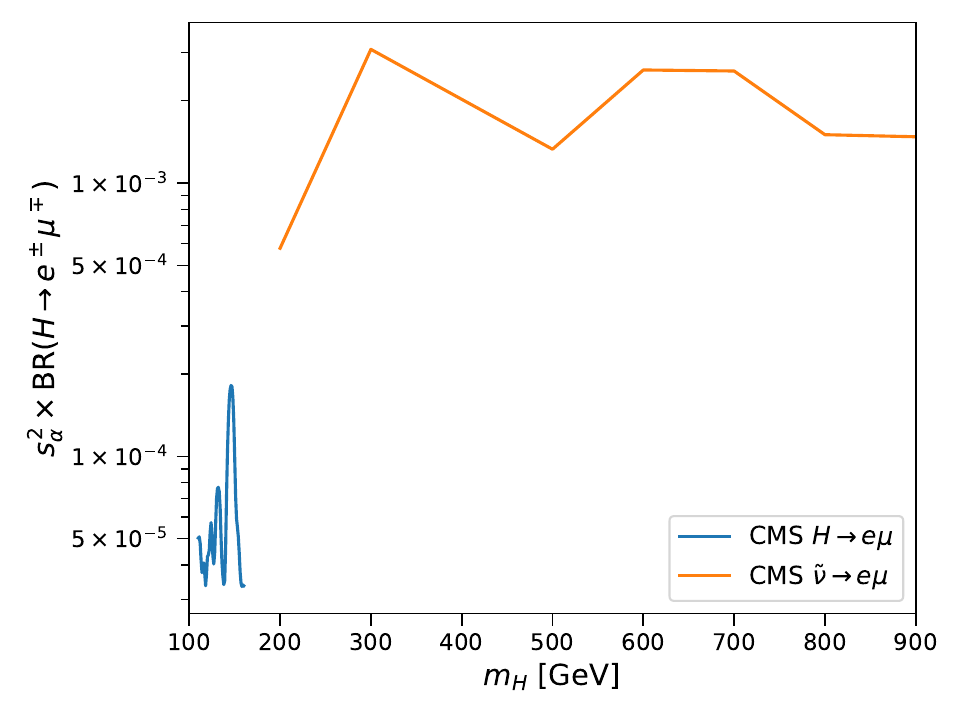}
\caption{Upper limits on $s^2_\alpha \times \text{BR}(H \rightarrow e^\pm\mu^\mp)$ from the single $H$ boson production. The constraints from the direct CMS search~\cite{CMS:2023pte} are shown in blue, while the limits from recasting the CMS sneutrino search~\cite{CMS:2022fsw} are shown in orange.}
\label{fig:H}
\end{center}
\end{figure}

Additional constraints arise from the pair production of $H$ and $A$ via an $s$-channel $Z$ boson, followed by $H \to A Z$ and $A \to e^\pm \mu^\mp$. The resulting final state contains four leptons, coming from the  $AA$ pair, and the $Z$ boson. The $Z$ boson will then decay to a pair of leptons, quarks or missing energy. This process is constrained by the CMS multilepton analysis~\cite{CMS:2021cox}, in particular the $4\ell$G and $4\ell$H signal regions. Both signal regions require at least four leptons with $p_T > 10$ GeV. If more leptons are present, the four highest $p_T$ leptons are selected. In the $4\ell$G signal region, the four selected leptons are grouped into pairs of opposite-sign same-flavor (OSSF) leptons. The two pairs are indicated by $Z_1$ and $Z_2$, where the invariant mass of $Z_1$ is the closest to the mass of the $Z$ boson. The region is further categorized according to the invariant mass of $Z_2$, and the transverse mass variable ($M_{T2}$) that involves $Z_1$, $Z_2$ and the missing transverse energy. Meanwhile, the signal region $4\ell$H requires exactly one OSSF pair, denote $Z_1$. The signal region is characterized by the invariant mass of $Z_1$ and the separation $\Delta R$ between the non-OSSF leptons pair. 

\begin{figure}
     \centering
     \subfigure[\label{fig:GZZ}]{
         \includegraphics[width=0.45\textwidth]{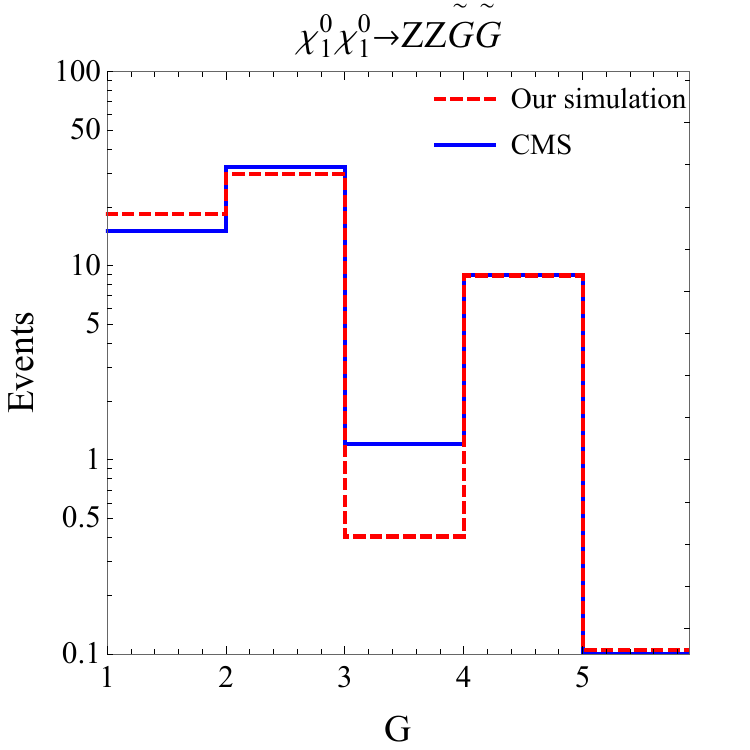}
         %\caption{Invariant mass of the four-lepton system, $m_{4\ell}$.}
         }
     %\hfill
     \subfigure[\label{fig:GHZ}]{
         \includegraphics[width=0.45\textwidth]{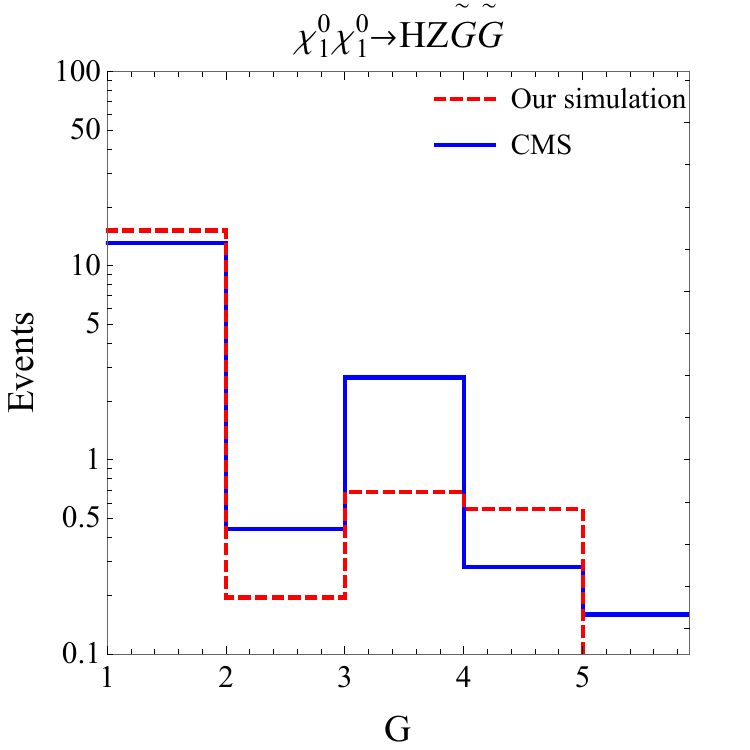}
         %\caption{Invariant mass of the OSDF pair, $m_{e^\pm\mu^\mp}$.}
         }
         \subfigure[\label{fig:GHH}]{
         \includegraphics[width=0.45\textwidth]{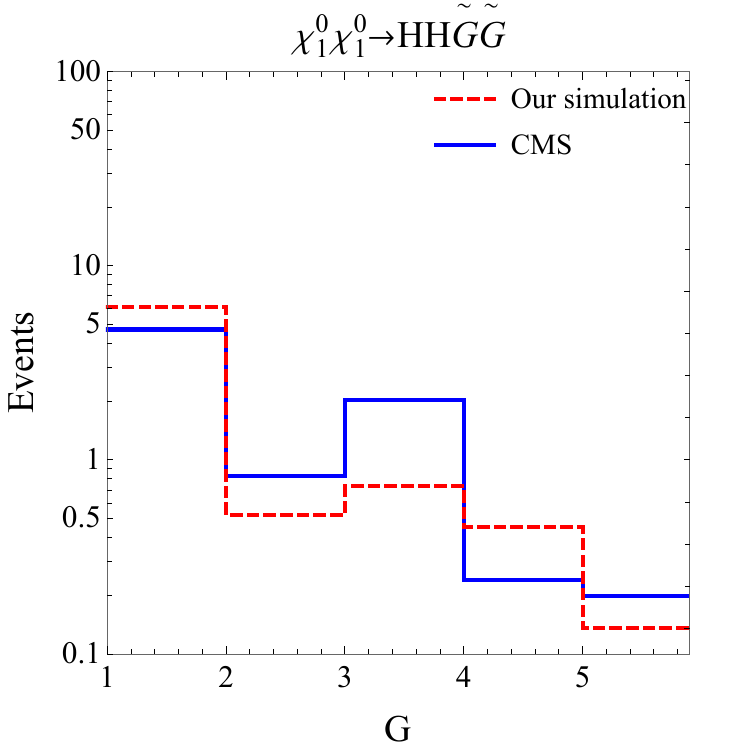}
         %\caption{Invariant mass of the OSDF pair, $m_{e^\pm\mu^\mp}$.}
         }
         \subfigure[\label{fig:Gmult}]{
         \includegraphics[width=0.45\textwidth]{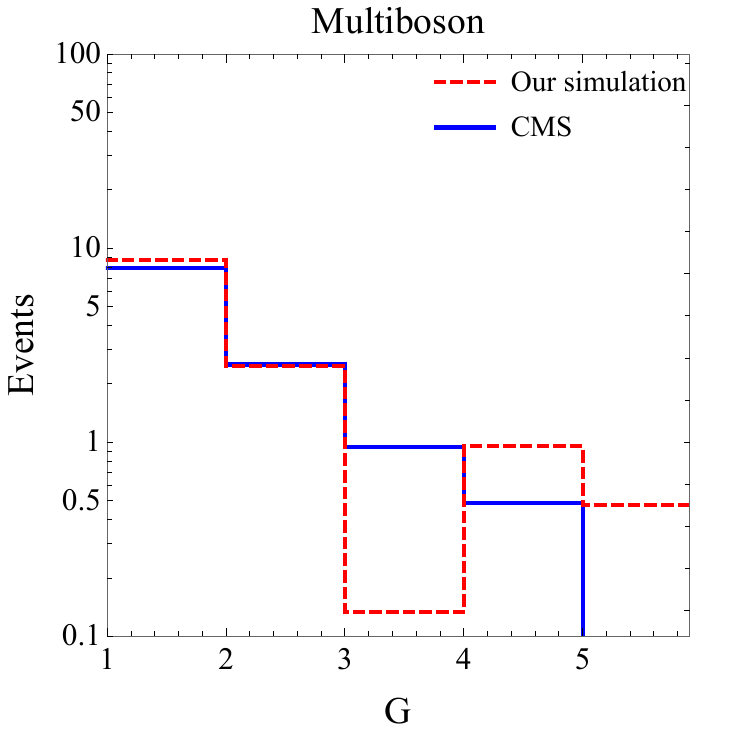}
         %\caption{Invariant mass of the OSDF pair, $m_{e^\pm\mu^\mp}$.}
         }
        \caption{The comparison between our simulations and CMS estimates for the 4$\ell$G signal regions.}
        \label{fig:G}
\end{figure}

\begin{figure}
     \centering
     \subfigure[\label{fig:HHZ}]{
         \includegraphics[width=0.45\textwidth]{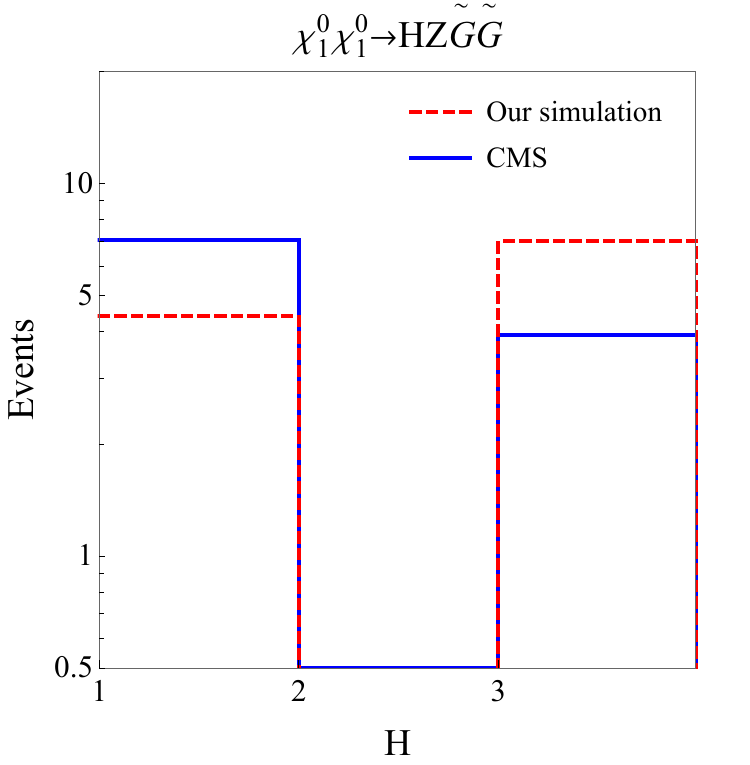}
         %\caption{Invariant mass of the OSDF pair, $m_{e^\pm\mu^\mp}$.}
         }
         \subfigure[\label{fig:HHH}]{
         \includegraphics[width=0.45\textwidth]{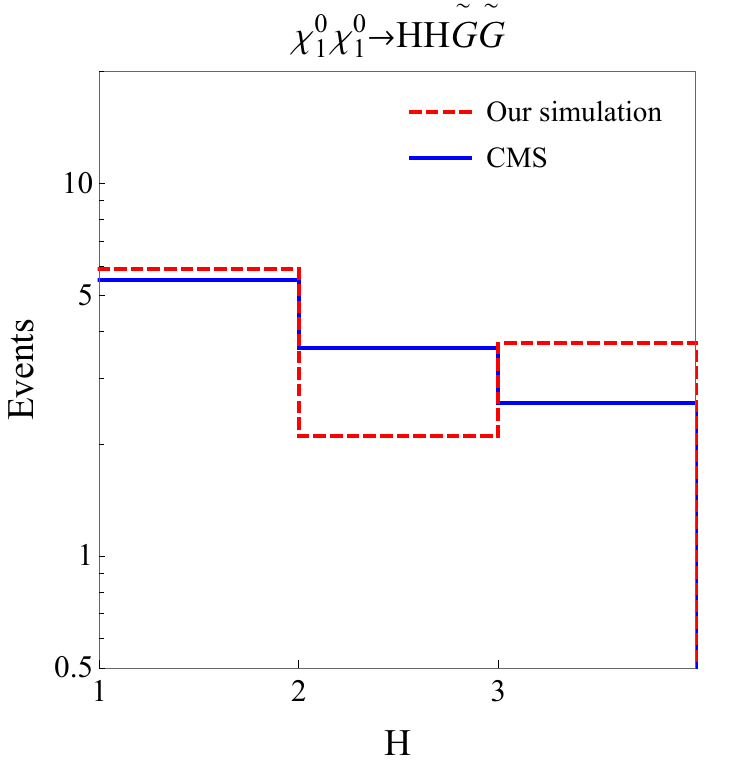}
         %\caption{Invariant mass of the OSDF pair, $m_{e^\pm\mu^\mp}$.}
         }
         \subfigure[\label{fig:Hmult}]{
         \includegraphics[width=0.45\textwidth]{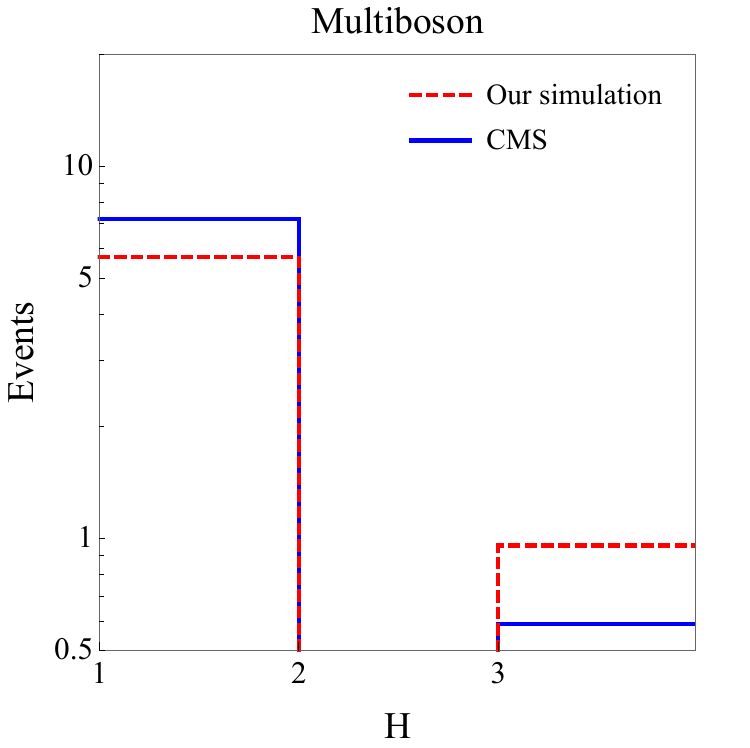}
         %\caption{Invariant mass of the OSDF pair, $m_{e^\pm\mu^\mp}$.}
         }
        \caption{The comparison between our simulations and CMS estimates~\cite{CMS:2021cox} for the 4$\ell$H signal regions.}
        \label{fig:H}
\end{figure}

To recast the CMS multilepton constraints, we use Feynrules2.3~\cite{Alloul:2013bka} to generate a UFO model file for the signal. The events are then simulated by Madgraph5~\cite{Alwall:2014hca} and passed through Pythia8~\cite{Bierlich:2022pfr} for parton showering and hadronization simulations. The detector responses are modeled using Delphes3~\cite{deFavereau:2013fsa} and the resulting events are analyzed using MadAnalysis5~\cite{Conte:2012fm}. %Finally, the CMS COMBINE tool~\cite{CMS:2024onh} is used to make statistical inference. %{\color{magenta}In particular, we employ \verb|HybridNew| method to model our statistics based on counting analysis, as it is more suitable for low-event case.}

%{\color{magenta}For the statistical analysis, we employ \verb|HybridNew| method to model our statistics based on counting analysis, as it is more suitable for low-event case. In the multilepton case, we divide the bins into 8 signal regions, while in the optimized case, it consists of one signal region. In both cases, the systematic uncertainties are incorporated into single nuisance parameter representing the integrated luminosity, assumed to be log-normal.}

As a check on our simulations, we compare our results against CMS estimate in figures~\ref{fig:G} and \ref{fig:H}. %A comparison between our simulations and the CMS estimates is presented in figures~\ref{fig:G} and \ref{fig:H}. 
Following CMS analysis, we simulate signal events coming from the production of neutralino pairs followed by decays to to various final states involving gravitinos: $\chi^0_1 \chi^0_1 \rightarrow ZZ\tilde G \tilde G$, $\chi^0_1 \chi^0_1 \rightarrow HZ\tilde G \tilde G$ and  $\chi^0_1 \chi^0_1 \rightarrow HH\tilde G \tilde G$. For the SM background, we simulate the multiboson process, as it is the only dominant background that contains neutrinos in the final states for the signal regions 4$\ell$G and 4$\ell$H.

%For statistical analysis, we use the CMS COMBINE tool~\cite{CMS:2024onh} with the \verb|HybridNew| method. 
Following CMS analysis, we divide the 4$\ell$G and 4$\ell$H categories into 5 and 3 signal regions, respectively. We then use the CMS COMBINE tool~\cite{CMS:2024onh} with the \verb|HybridNew| method to extract the upper limit on the $A\to e^\mp\mu^\pm$ branching fraction. Specifically, we model our statistical analysis as a counting experiment with 8 different signal regions. We have also introduced a nuisance parameter, whose distribution is assumed to be log-normal, to account for systematic uncertainties.

\begin{figure}[t]
\begin{center}
\includegraphics[width=0.5\textwidth]{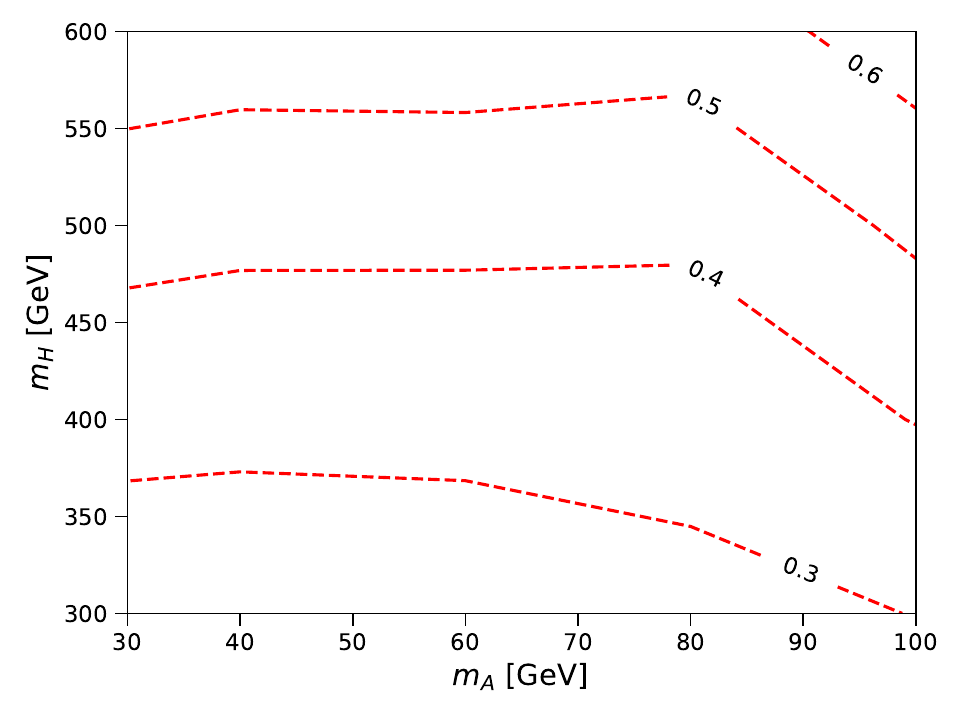}
\caption{The contours of upper bounds on the branching ratio $\mathrm{BR}(A \to e^\pm\mu^\mp)$ from the $H$ and $A$ pair production.}
\label{fig:HAbound}
\end{center}
\end{figure}

From our analysis we find that if $A$ decays only to $e^\pm\mu^\mp$ pairs, the multilepton search imposes a stringent constraint on the mass of $H$, i.e., $m_H \gtrsim 850$ GeV. Such a heavy $m_H$ is in serious tension with perturbativity and vacuum stability requirements. The bound can be relaxed if $A$ admits an additional decay mode that is less visible, such as $A\to \tau^+\tau^-$, which dilutes the branching fraction of $A\to e^\pm\mu^\mp$. For instance, if the branching fraction BR$(H \rightarrow e^\pm\mu^\mp)$ is reduced to $\sim 50\%$, a relatively light $m_H=600$ GeV consistent with the multilepton bounds can be obtained.  Figure~\ref{fig:HAbound} shows a contour of the upper limit on BR$(H \rightarrow e^\pm\mu^\mp)$ as a function of $m_A$ and $m_H$. For the remaining of this paper, we choose the value of BR$(A \rightarrow e^\pm\mu^\mp)$ to be 50\% as a benchmark.

%%%%%%%%%%%%%%%%%%%%%%%%%%%%%%%%%%%%%%%%
\section{Novel searches for LFV from the SM-like Higgs decay}
\label{sec:collider}
%%%%%%%%%%%%%%%%%%%%%%%%%%%%%%%%%%%%%%%%
In this section, we propose two new LFV signatures of the 125-GeV Higgs boson $h$. For a sufficiently light pseudoscalar $A$, the decays $h\to AZ^{(\ast)}$ and $h\to AA$ are open. The two decay modes can lead to 4-lepton final states and can be constrained by the existing data.

%%%%%%%%%%%%%%%%%%%%%%%%%%%%%%%%%%%%%%%%%
\subsection{$h \rightarrow A Z$}
\label{sec:hAZ}

\begin{figure}[t]
\begin{center}
\includegraphics[width=0.5\textwidth]{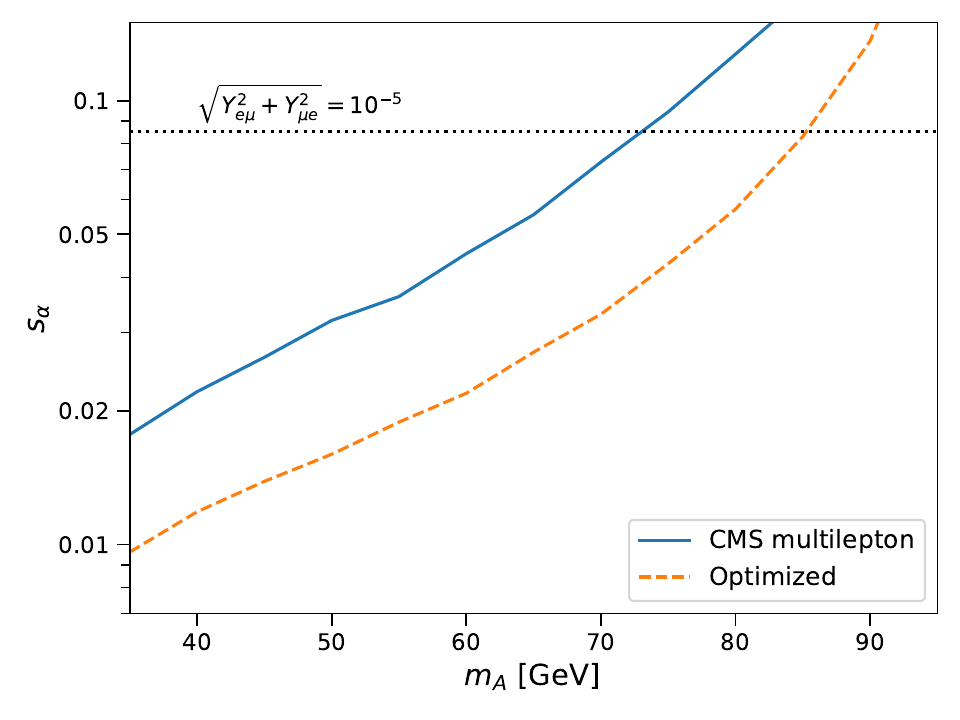}
\caption{The upper bounds on $\sin \alpha$ derived from the CMS multilepton searches~\cite{CMS:2021cox} assuming BR$(A \rightarrow e^\pm\mu^\mp) = 50\%$ (solid blue) and the optimized limits which require the invariant mass of the four-lepton to lie within $m_h\pm2.5$ GeV range (dashed orange). For comparison, the corresponding constraints from low energy LFV search, assuming $m_H=m_{H^\pm}=600$~GeV, are also given (dotted black).}
\label{fig:hAZ}
\end{center}
\end{figure}

We first discuss the decay channel $h \to A Z^{(\ast)}$. Such a decay process is possible if the mixing angle $s_\alpha$ is non-vanishing. The $h \to A Z^{(\ast)}$ can lead to a four-lepton final state with an OSSF pair from $Z^{(\ast)}$ and an opposite-sign different-flavor (OSDF) pair from the $A$. This signature is covered by the $4\ell$H signal region in the CMS multilepton analysis~\cite{CMS:2021cox}. By recasting the CMS multilepton bounds, as discussed in the previous section, we extract limits on the mixing angle $s_\alpha$. Assuming the branching ratio BR$(A \rightarrow e^\pm\mu^\mp) = 50\%$, the corresponding limits on $s_\alpha$ are shown as the solid blue line in figure~\ref{fig:hAZ}. We note that for this branching fraction assumption, the multilepton constraints from $HA$ production require $m_H \gtrsim 550$ GeV, as shown in figure~\ref{fig:HAbound}.

\begin{figure}
     \centering
     \subfigure[Invariant mass of the four-lepton system, $m_{4\ell}$.\label{fig:m4l}]{
         \includegraphics[width=0.45\textwidth]{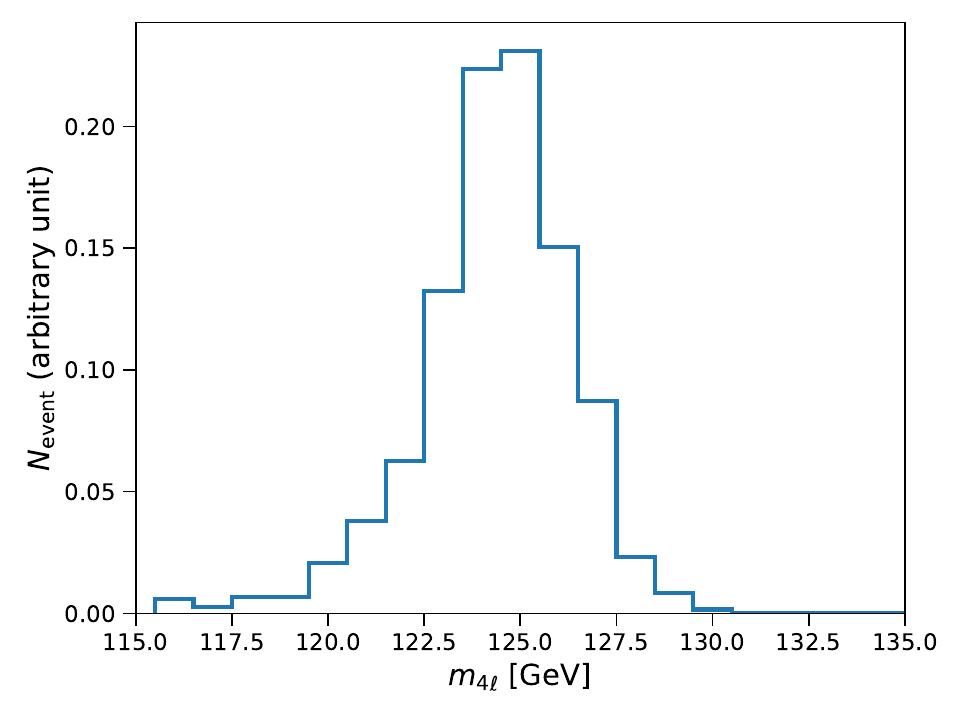}
         %\caption{Invariant mass of the four-lepton system, $m_{4\ell}$.}
         }
     %\hfill
     \subfigure[Invariant mass of the OSDF pair, $m_{e^\pm\mu^\mp}$.\label{fig:memu}]{
         \includegraphics[width=0.45\textwidth]{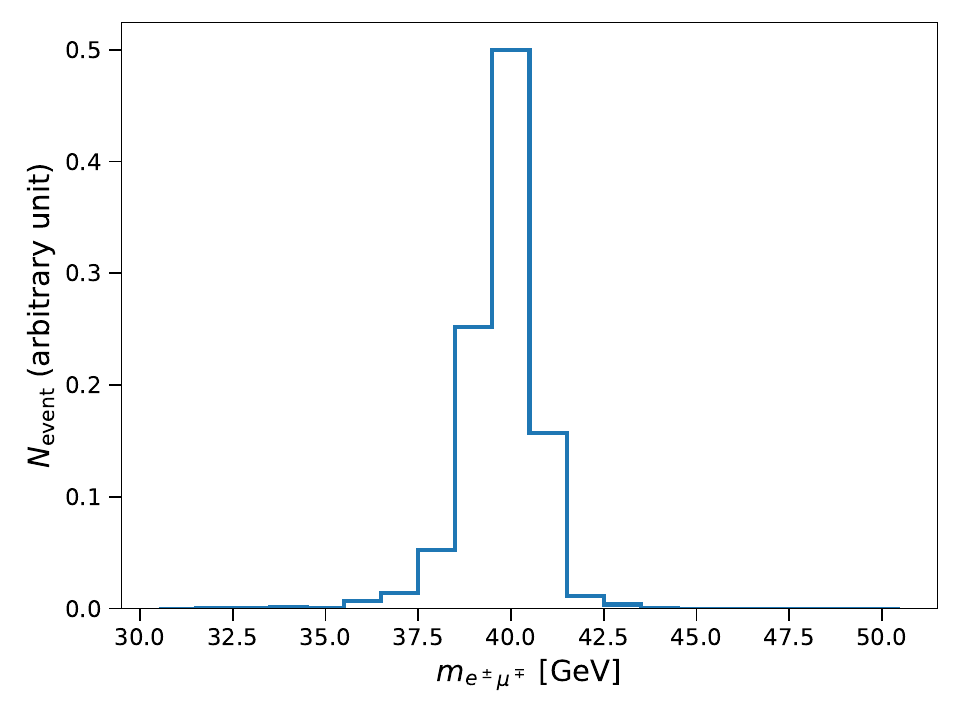}
         %\caption{Invariant mass of the OSDF pair, $m_{e^\pm\mu^\mp}$.}
         }
        \caption{The simulated kinematic distributions for the $h \rightarrow AZ \rightarrow e^\pm\mu^\mp \ell^+\ell^-$ decay process assuming $m_A = 40$ GeV.}
        \label{fig:massdist}
\end{figure}

The bounds can be further improved by exploiting the kinematics features of the decay $h\to AZ^{(*)}\to4\ell$. The invariant mass of the four leptons, $m_{4\ell}$, peaks near $m_h$ as shown in figure~\ref{fig:m4l}. Furthermore, the invariant mass of one of the two possible OSDF pair, $e^\pm\mu^\mp$, would reproduce the pseudocalar mass $m_A$. The distribution of the invariant mass of the OSDF pair is shown in figure~\ref{fig:memu}.

Motivated by these features, we define an optimized signal region as follows. The events must contain four leptons with an OSSF pair and an OSDF pair ($\ell^+\ell^- e^\pm\mu^{\mp}$). The four-lepton invariant mass must lie within a 5-GeV window around $m_h$, i.e., $|m_{4\ell} - m_h| < 2.5$ GeV. Furthermore, at least one of the OSDF pairs must have invariant mass within a 5~GeV window of $m_A$, $|m_{e^\pm\mu^\mp}-m_A| < 2.5$ GeV. Moreover, to suppress the background involving top quarks, events containing $b$-jets are vetoed.

\begin{figure}[t]
\begin{center}
\includegraphics[width=0.5\textwidth]{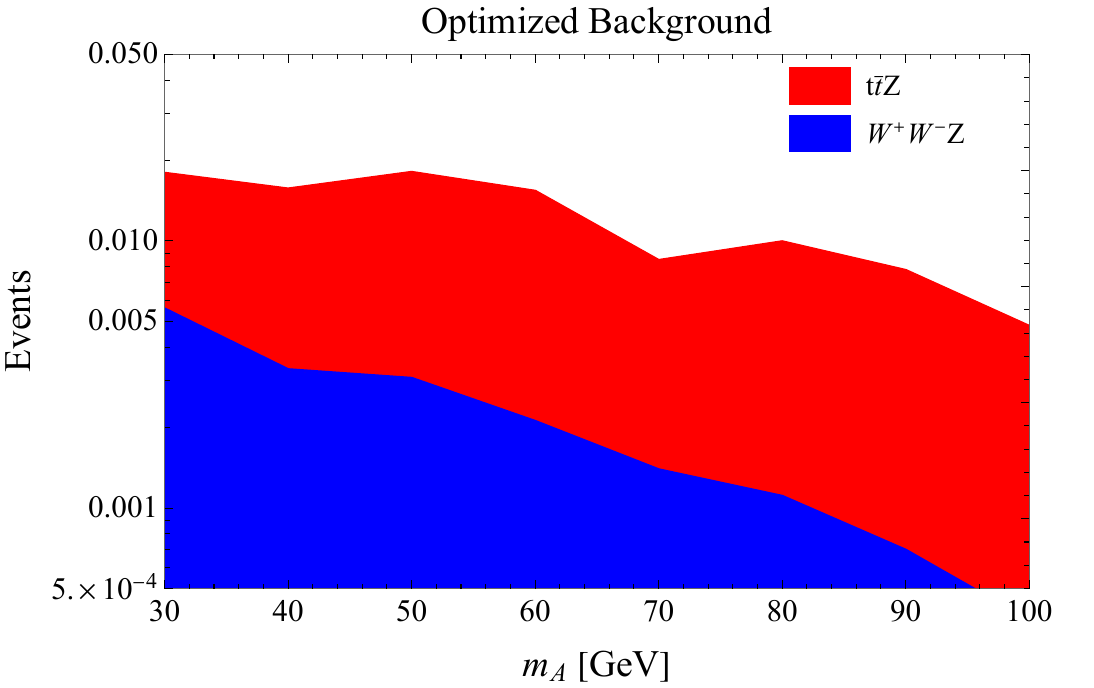}
\caption{The number of expected background for the optimized signal region as a function of the pseudoscalar mass, $m_A$.}
\label{fig:hAZback}
\end{center}
\end{figure}

The dominant backgrounds for our optimized signal region are the multiboson productions--- $W^+ W^- Z$ and $2W^+ 2W^-$; and the top-associated processes--- $t\bar t Z$, $t\bar t t \bar t$ and $t\bar t W^+ W^-$. The multiboson backgrounds are expected to be small due to the small leptonic partial decay widths of $W$ and $Z$. On the other hand, the top-associated backgrounds are suppressed by the $b$-jet veto. {To simulate the background processes, we employed Madgraph5~\cite{Alwall:2014hca}, Pythia8~\cite{Bierlich:2022pfr} and Delphes3~\cite{deFavereau:2013fsa}. 
%The Delphes card used for the simulation the the identical to the CMS card adopted to reproduce the mutilepton signatures described in Section \Ref{sec:colliderbound}. 
The Delphes card used for the detector simulation is identical to the one adopted for the mutilepton analysis described in section~\Ref{sec:colliderbound}.
The resulting distributions for the dominant background contributions are shown in figure~\ref{fig:hAZback}. Although a dedicated simulation of backgrounds arising from the mistagged leptons is beyond the scope of this work, our rough estimate indicate that they contribute less than 0.5 events,
%we provide a rough estimate indicating that the expected contribution from such events is below 0.5 events, 
assuming a jet-to-lepton fake rate of less than 1\%. %From our analysis, we estimate fewer than 0.5 background events in our optimized signal region across the pseudoscalar mass range 30 GeV $\leq m_A\leq$~100 GeV. The projected sensitivity for the optimized signal region is shown by the orange dashed line in figure~\ref{fig:hAZ}.
 The optimized upper limit is extracted using the CMS COMBINE tool~\cite{CMS:2024onh} with the \verb|HybridNew| method. For our optimized scenario, we model our statistics as a single bin counting experiment. In this estimate, we neglect systematic uncertainty.} %Again, as in the multilepton analysis in the previos section, we model systematic uncertainty with a log-normal nuisance parameter. 
%%%%%%%%%%%%%%%%%%%%%%%%%%%%%%%%%%%%%%%%%
\subsection{$h \rightarrow A A$}
\label{subsec:haa}
For sufficiently light pseudoscalar, $2m_A\le m_h$, the SM-like Higgs boson may also decay to a pair of pseudoscalars. This decay mode is controlled by the scalar quartic couplings $\lambda_3$, $\lambda_4$ and $\lambda_5$ with decay width given by 
\begin{equation}
    \Gamma_{h \to A A} = \frac{\lambda_{hAA}^2 v^2}{32 \pi m_h} \sqrt{1-\frac{4 m_A^2}{m_h^2}}
\end{equation}
where $\lambda_{hAA} \equiv \left(\lambda_3 + \lambda_4 - \lambda_5 \right)  c_\alpha$. The combination $\lambda_{hAA}$ appears in the mass relations, equations.~\eqref{eq:mHp}  and \eqref{eq:mA}. In the case where each of $A$ subsequently decays via $A \rightarrow e^\pm\mu^\mp$, the final state consists of two electrons and two muons ($2e 2\mu$). 

While existing searches for exotic Higgs decays to light pseudoscalar~\cite{ATLAS:2021ldb, CMS:2021pcy} primarily focus on flavor-conserving final states, they can still offer sensitivity to LFV scenarios. These analyses typically target configurations with two OSSF pairs ($4e$, $4\mu$ or $2e2\mu$), with the signal regions targeting the invariant mass equality of the two pairs. However, in our scenario with the LFV decays, the invariant masses of the two OSSF lepton pairs will generically differ, as they originate from different parent particles. Nonetheless, a partial overlap with the defined signal regions allows constraints to be derived.

\begin{figure}[t]
\begin{center}
\includegraphics[width=0.5\textwidth]{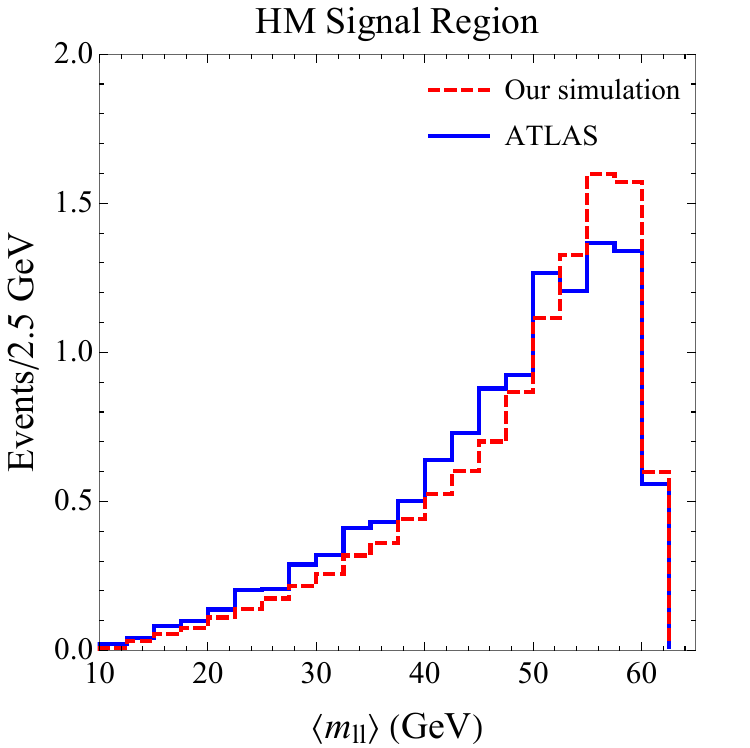}
\caption{The comparison between our simulations and ATLAS estimates~\cite{ATLAS:2021ldb} for the $H \rightarrow ZZ^* \rightarrow 4 \ell$ backgrounds.}
\label{fig:AtlasBackground}
\end{center}
\end{figure}

In this work, we will recast the ATLAS search~\cite{ATLAS:2021ldb}, focusing on the high-mass signal region, which requires at least four light leptons in the `Loose' category. The baseline electrons (muons) are required to have $p_T > 7(5)$ GeV and $|\eta| < 2.47(2.5)$. The events must have two OSSF pairs with the three leading leptons satisfying $p_T >$ 20 GeV, 15 GeV and 10 GeV. Two OSSF pairs are then formed and labeled by their invariant masses $m_{12}$ and $m_{34}$, where $m_{12}$ is the pair closest to the mass of the $Z$ boson. For the $4e$ and $4\mu$ final states, the pairing that minimizes the difference $|m_{12}-m_{34}|$ is chosen. Since the signal comes from the SM-like Higgs decay, events are retained if the four-leptons invariant mass falls within 115 GeV $< m_{4\ell} < $ 130 GeV. Moreover a $Z$-veto is applied by imposing 10 GeV $< m_{12,34} <$ 64 GeV. For the $4e$ and $4\mu$ channel, an additional condition 5 GeV $< m_{14,23} <$ 75 GeV is also required, where $m_{14}$ and $m_{23}$ represent the other combination of OSSF pair. Finally, a ratio cut is imposed on the OSSF invariant masses, requiring $m_{34}/m_{12} > 0.85 - 0.1125 f(m_{12})$, where the function $f(x)$ is given in the appendix of ref.~\cite{ATLAS:2021ldb}. The function $f(x)$ is chosen so that the ratio of $m_{34}/m_{12}$ is close to one, while taking into account the detector performance. The results are then shown in bins of $\langle m_{\ell\ell}\rangle = \frac{1}{2} \left( m_{12}+m_{34} \right)$. From the data, ATLAS observes a total of 20 events, consistent with the SM expectation. {The comparison between our simulations and the ATLAS estimates is presented in figure~\ref{fig:AtlasBackground}. To obtain these results, we modified the default Delphes card for ATLAS to incorporate the selection efficiencies of Loose electrons~\cite{ATLAS:2019qmc} and muons~\cite{ATLAS:2020auj}.}

\begin{figure}[t]
\begin{center}
\includegraphics[width=0.5\textwidth]{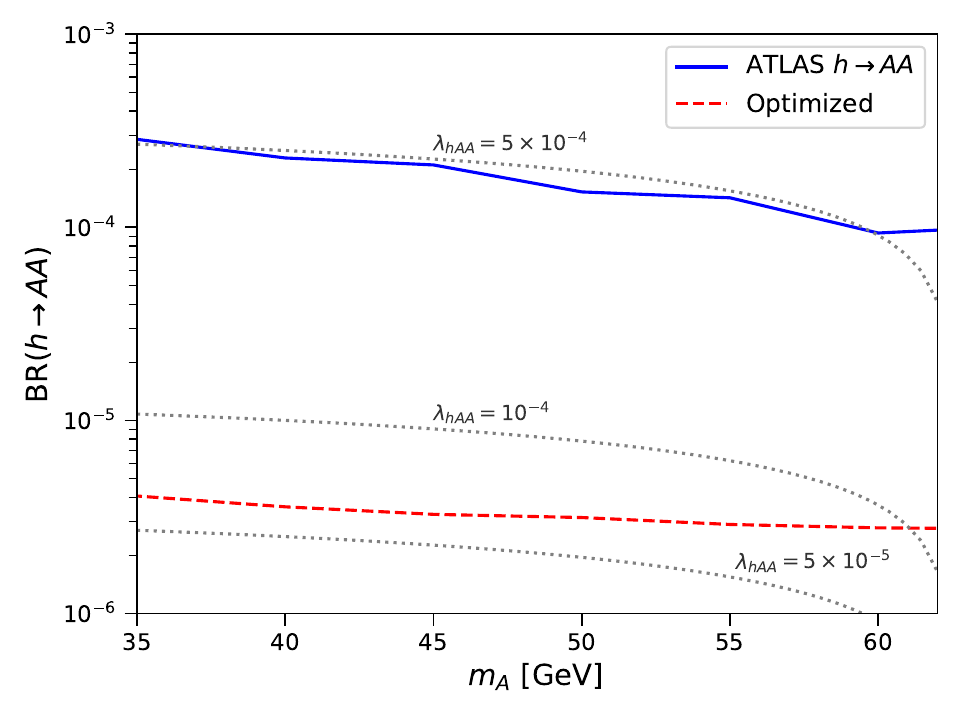}
\caption{The upper bounds on the branching fraction BR$(h \rightarrow AA)$, assuming the BR$(A \rightarrow e^\pm\mu^\mp) = 50\%$. For comparison, the branching ratios BR$(h \rightarrow AA)$ as a function of quartic coupling $\lambda_{hAA}\equiv(\lambda_3+\lambda_4+\lambda_5)c_\alpha$ are shown by the dotted lines for $\lambda_{hAA}=5\times10^{-14}$, $10^{-4}$ and $5\times10^{-5}$.}
\label{fig:hAA}
\end{center}
\end{figure}

To estimate the LFV sensitivity, we simulate the $h \rightarrow A A \rightarrow e^+\mu^- e^-\mu^+$ signal using the same simulation chain as in the multilepton analysis. We find approximately $30\%$ of the events that pass pre-selection cuts satisfy the ratio $m_{34}/{m_{12}} > 0.85 - 0.1125 f(m_{12})$ criterion. The resulting upper bound on the branching ratio BR$(h\to AA)$, assuming BR$(A\to e^\pm\mu^\mp) = 50\%$, is shown as the solid blue line in figure~\ref{fig:hAA}. This combination already puts a strong constraint on the couplings combination $\lambda_{hAA}$.

To enhance the sensitivity of the LFV signature, we propose an alternative signal region that focuses on the final states with a pair of same-sign electrons and a pair of same-sign muons. The baseline lepton identification and the transverse momentum thresholds remain the same as those used in the ATLAS analysis. The four-lepton invariant mass is again required to lie within the Higgs mass window, between 115 GeV and 130 GeV. In this configuration, potential backgrounds are extremely suppressed. Processes such as $hh$ and $WWWW$ productions could, in principle, contribute to the backgrounds. However, their cross sections and leptonic branching fractions are sufficiently small to render the backgrounds negligible. {At the partonic level, the SM processes producing a pair of same-sign electrons and a pair of same-sign muons accompanied by four (anti)neutrinos, within the invariant mass window 115 GeV $< m_{4\ell} < 130$ GeV, have cross sections of order  $10^{-5}$ fb, resulting in 0.01 events. We further estimate that the background arising from misidentified leptons contributes fewer than 0.1 events, provided that the mistag rate is below 1\%. Finally, the upper limit on the branching fraction $h\to AA$ is extracted by the same statistical procedure as in the $h\to AZ$ analysis.}
The projected upper limit on BR$(h\to AA)$ under this optimized signal region is shown as the red dashed line in figure~\ref{fig:hAA}.

%%%%%%%%%%%%%%%%%%%%%%%%%%%%%%%%%%%%%%%%%
\section{Conclusion and discussion}
\label{sec:conc}

We have investigated electron–muon flavor-violating Yukawa couplings of the SM-like Higgs boson within the type-III 2HDM, 
focusing on a scenario in which the pseudoscalar is lighter than the 125-GeV Higgs boson, with a sizable branching fraction $A\to e^\pm\mu^\mp$. In this framework, we propose two novel decay signatures of the SM-like Higgs boson: $h\to AZ^{(*)}\to e^\pm\mu^\mp\ell^+\ell^-$  and $h\to AA\to 2e^\pm2\mu^\mp$. The former channel is constrained by the existing CMS multilepton analysis~\cite{CMS:2021cox}, resulting in the bounds on the mixing angle $s_\alpha$ between the two neutral CP even Higgs bosons, see figure~\ref{fig:hAZ}. The latter process is constrained by the ATLAS $h\to AA$ analysis~\cite{ATLAS:2021ldb} that leads to upper limits on the branching fraction BR$(h\to AA)$, see figure~\ref{fig:hAA}. 

The constraints on $s_\alpha$ and BR$(h\to AA)$ above can be further improved by taking advantage of the kinematics of the decay processes. By requiring the four-lepton invariant mass lies close to $m_h$ and the invariant mass of the OSDF pair lies close to $m_A$, the backgrounds for $h\to AZ^{(*)}\to e^\pm\mu^\mp\ell^+\ell^-$ and $h\to AA\to 2e^\pm2\mu^\mp$ become negligible. Using these optimized signal regions, the limit on $s_\alpha$ improves roughly by a factor of two for the $h\to AZ^{(*)}\to e^\pm\mu^\mp\ell^+\ell^-$ signatures, see figure~\ref{fig:hAZ}. In the case of $h\to AA\to 2e^\pm2\mu^\mp$ signatures, the limit on the branching fraction $h\to AA$ improves by an order of magnitude, see figure~\ref{fig:hAA}.

Beyond these Higgs channels, additional constraints arise from the direct search for a resonance decaying to a pair of $e\mu$. In our scenario, only the heavy Higgs $H$ can be singly produced at the LHC. The bound from the direct search $H\to e\mu$ is relevant only for $m_H < m_A + m_Z$, where the dominant decay mode $H\to AZ$ is not accessible. When the decay channel $H\to AZ$ is open, the constraints on the model come mainly from the pair production of $H$ and $A$, following by the subsequent decays $H\to AZ$ and $A\to e\mu$. As a result, the final state of this production mechanism consists of a pair of electrons, a pair of muons and a $Z$ boson. This scenario is tightly constrained by the CMS multilepton search, pushing the $m_H > 500$ GeV for the branching ratio BR($A \rightarrow e^\pm\mu^\mp)> 50\%$). We have also considered constraints from low-energy LFV searches $\mu\to e\gamma$ and $\mu\to e$ conversion, which place upper limits on the combination $s_{2\alpha}\sqrt{Y_{e\mu}^2+Y_{\mu e}^2}$, see figure~\ref{fig:lfvconstraint}. These low-energy LFV processes offer complementary constraints to our novel collider searches for LFV in the decays of the Higgs bosons. 

In conclusion, we have investigated various constraints on the 2HDM type-III with a light pseudocalar and electron-muon flavor violating Yukawa couplings. In addition to the existing limits of multilepton searches, resonance searches and low-energy flavor observables, we have identified two novel SM-like Higgs decay signatures, $h\to AZ^{(*)}\to e^\pm\mu^\mp \ell^+\ell^-$ and $h\to AA\to 2e^\pm2\mu^\mp$. We have shown that once the optimized kinematic selections are used, the sensitivity of the LHC to these channels  improves substantially, with the limits of $s_\alpha$ strengthened by a factor of two and the constraints of $\mathrm{BR}(h\to AA)$ enhanced by an order of magnitude. These new probes help cover the parameter space, particularly in regions where conventional searches lose sensitivity. %These new probes provide complementary coverage of the parameter space, particularly in regions where conventional searches lose sensitivity. 

%%%%%%%%%%%%%%%%%%%%%%%%%%%%%%%%%%%%%%%%
\acknowledgments{We would like to thank M. Sher for useful discussion. 
R.P. was supported by Direktorat Riset, Teknologi dan Pengabdian kepada Masyarakat Direktorat Jenderal Pendidikan Tinggi, Riset, dan Teknologi, Kementerian Pendidikan, Kebudayaan, Riset, dan Teknologi
Republik Indonesia in the year 2024 with contract number III/LPPM/2024-06/110-PE and 006/SP2H/\linebreak RTMONO/LL4/2024 and Direktorat Penelitian dan Pengabdian kepada Masyarakat, Direktorat Jenderal Riset dan Pengembangan, Kementerian Pendidikan Tinggi, Sains dan Teknologi
Republik Indonesia in the year 2025 with contract number 7939/LL4/PG/2025; III/LPPM/ 2025-06/154-PE and 125/C3/DT.05.00/PL/2025. The work of P.U. was supported in part by Thailand NSRF via PMU-B under grant number B39G680009. 
The authors also acknowledge the National Science and Technology Development Agency, National e-Science Infrastructure Consortium, Chulalongkorn University and the Chulalongkorn Academic Advancement into Its 2nd Century Project, NSRF via the Program Management Unit for Human Resources and Institutional Development, Research and Innovation [Grant No.B39G680009] (Thailand) for providing computing infrastructure that has contributed to the research results reported within this paper.}

%%%%%%%%%%%%%%%%%%%%%%%%%%%%%%%%%%%%%%%%%
\bibliography{reference} 
\bibliographystyle{JHEP}
%%%%%%%%%%%%%%%%%%%%%%%%%%%%%%%%%%%%%%%%%
\end{document}